# Optoelectronic characterisation of twisted germanium sulfide nanowires with experimental observation of intrinsic ferroelectricity

*Tamaghna Chowdhury[1], Aastha Vasdev[2], Goutam Sheet[2] and Atikur Rahman[1]\**

[1]Department of Physics, Indian Institute of Science Education and Research (IISER), Pune, Maharashtra, 411008, India.
[2]Department of Physical Sciences, Indian Institute of Science Education and Research Mohali, Knowledge City, Sector 81, Mohali 140306, India.

Abstract

We report the optoelectrical characterisation of Eshelby twisted Germanium sulfide (GeS) nanowires with first experimental observation of ferroelectric order at room temperature in GeS which is an otherwise centrosymmetric molecule. The chemical composition and structure of these nanowires were confirmed by various spectroscopic, microscopic and diffractive techniques. In addition, the nanowires were found to be stable over time. From the optoelectronic measurements we found that these p-type semiconducting GeS nanowires have up to two orders higher charge carrier mobility than GeS nanosheets and have sustainable, robust photo-switching property. The existence of room temperature ferroelectricity is confirmed by piezoresponse force microscopy which showed hysteresis and butterfly loop, characteristics of a ferroelectric material. Our observations reveal that the properties of twisted GeS nanowires can be harnessed in making efficient electric sensors, photodetectors, data memories and flexible electronics.

Keywords

Twisted GeS nanowires; VLS growth; ferroelectric; PFM; transition metal dichalcogenides

Introduction

After the successful isolation of graphene in 2004 by Novoselov et.al.,[1] an wide range of transition metal dichalcogenides (TMDC's) have been discovered because of their fundamental interests and potential applications. These TMDC's offer unique electrical, thermal, mechanical and optical properties which were otherwise absent in their 3D counterparts. Among the TMDC's , those having piezoelectric and ferroelectric properties have grown considerable interests in recent years. Ferroelectricity is a property of spontaneous electric polarization that can be controlled by an external electric field.[2,3] These TMDC based ferroelectrics have advantages like band gap tunability and flexibility, promising its applications as electric sensors and data memories.[2,3] Another growing sector in 2D research area is twistronics.[4] Twistronics has attracted considerable attention in recent times because of its implications in fundamental physics (e.g the discovery of superconductivity in twisted bilayer graphene[5]) as well as its potential applications in electronics industry. Twist is a new degree of free- dom through which we can modify various electrical and optical property of van der Waals heterostructures. [6]

A plethora of TMDC's have been experimentally confirmed or theoretically predicted as ferroelectric, like monolayer TMDCs (such as $MoS_2$, $MoSe_2$, $WS_2$ and $WSe_2$)[2,3] and group- IV monochalcogenides (such as GeS, GeSe and SnS).[2,3,7] Traditionally, non-centrosymmetric materials are expected to show ferroelectricity but GeS is a centrosymmetric molecule. However, if one can break this centrosymmetry by any means then GeS might exhibit ferroelectricity, which has been theoretically predicted.[7–9] We wanted to use twist in GeS as a "tool" for breaking centrosymmetry, but main challenge lies in the huge difficulty in preparation of GeS using exfoliation or CVD methods and so its application as a ferroelectric material is yet to be experimentally verified. Another challenging task lies in the growth of twisted GeS nanostructures using transfer stacking method is that one doesn't really have much control over the twist angle and the layers often relaxes making it a highly complex and less reliable method. Also to get a good quality sample, keeping the interface clean, one has to take care of multiple parameters which are very difficult to control. Recently, Liu, Y. et al. and Sutter, P. et al. reported a robust growth process of intrinsically twisted GeS nanowires.[10,11] In this paper we report, successful growth of Eshelby twisted stable GeS nanowires and because of this twist, inversion symmetry is spontaneously broken in GeS and hence it becomes intrinsically ferroelectric. In 2D ferroelectrics there is a significant reduction of in-plane ferroelectricity by size reduction (as in ribbons or in flakes) which is not the case in one dimensional ferroelectric materials, so we chose to work with GeS nanowires formed by layer of 2D twisted layers instead

of mono-layers.[7]

The synthesis of Eshelby twisted GeS nanowires were done using low pressure chemical vapor deposition (LPCVD) method. Chemical composition and phase of these nanowires are thoroughly verified by PXRD, SAED, EDS, TEM and Raman scattering. The twist in these nanowires can be easily seen in field emission scanning electron microscope (FE-SEM). A crystallographic twist is also evident from selected area electron diffraction (SAED) pattern. The optoelectronic characteristics of these nanowires were studied by various electronic and optical measurements which showed these nanowires have good photoresponse and photo-switching property. The ferroelectric nature of these nanowires was studied by piezoresponse force microscopy (PFM). The PFM further confirms the presence of spontaneous polarization and $180°$ polarization switching under an external field at room temperature.

Results and discussion

The twisted GeS nanowires were grown by vapor-liquid-solid (VLS) technique.[11] In VLS technique gold acts as a catalyst, for this 3 nm of gold (Au) was deposited over a silicon (Si) substrate by thermal evaporation.[11] These Au coated substrates were put inside a quartz tube in a LPCVD setup along with 30 mg of GeS powder in an alumina boat. The precursor boat containing GeS powder was kept at a temperature of $450°C$ and the substrate boat was kept at a temperature of $270 - 300°C$. Ar balanced with 2% $H_2$ was used a carrier gas with a flow rate of 50 sccm.[10,11] Pressure is a very crucial parameter for the growth of these twisted nanowires, so we kept the quartz tube at a pressure of 1-2 torr for an optimal coverage of twisted nanowires over the substrate.[10] The whole system was kept at this condition for 30 minutes. The substrates were then taken out of quartz tube and we observed forest of nanowires covering the whole substrate under an optical microscope. Field emmission scanning electron microscope (FE-SEM) and we observed the twisted nanowires with gold at the tip (fig. 1a). Also the crystallographic twist is confirmed by TEM and SAED (fig.1b-d).

To confirm the chemical composition and to know the phase of the nanowires we performed powder X-Ray diffraction (PXRD), selected area electron diffraction pattern (SAED), Transmission electron microscopy (TEM), Raman scattering, and energy dispersive X-Ray analysis (EDS) experiments on the nanowires. The peaks obtained from the PXRD (fig.2a)

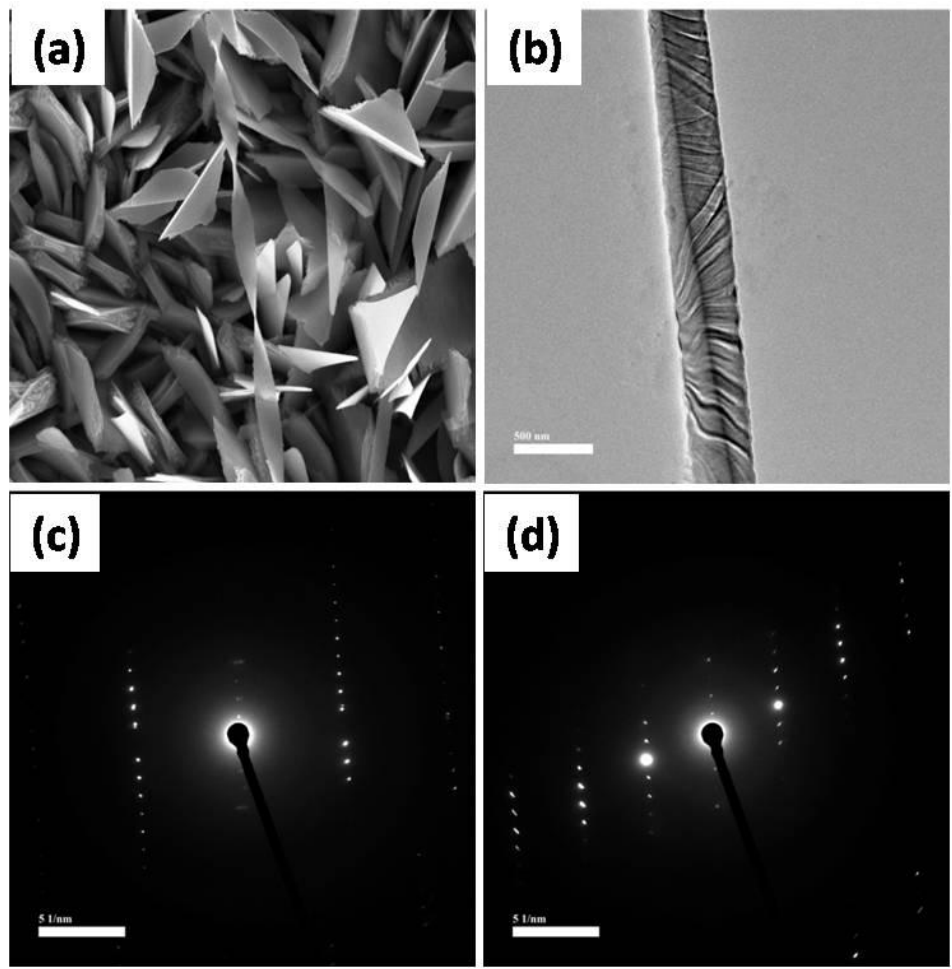

Figure 1: GeS nanowires with Eshelby twist. (a) FESEM image of a twisted GeS nanowire adhering to the substrate. (b) TEM image of a twisted GeS nanowire with a line defect at the centre of the wire. (c) and (d) SAED patterns taken from two different locations on the nanowire, suggesting that the crystal orientation changes with respect to the incident electron beam

on these nanowires exactly corresponds to the already reported PXRD peaks of GeS.[12,13] The GeS crystal belongs to the orthorhombic Pbnm space group (JCPDS No. 00-051-1168) with lattice constants of a = 4.30 Å, b = 10.47 Å, and c = 3.64 Å[12] . The existence of the Eshelby twist was confirmed by TEM studies. In TEM image (fig.1b) a line defect is evident at the centre of the nanowire. A crystallographic twist is also evident from selected area electron diffraction pattern (SAED) which shows (fig.1c and 1d) change in crystallographic direction along the length of the wire.[10] In the Raman scattering experiment (using 633 nm wavelength laser) phonon modes of the GeS nanowires were measured (fig. 2b), and the identified peaks were $A_g^4$ (270 cm$^{-1}$), $B_{1g}^2$ (212 cm$^{-1}$), $A_g^3$ (240 cm$^{-1}$) and $A_g^2$ (115 cm$^{-1}$) which are already

reported for GeS.[13–15]

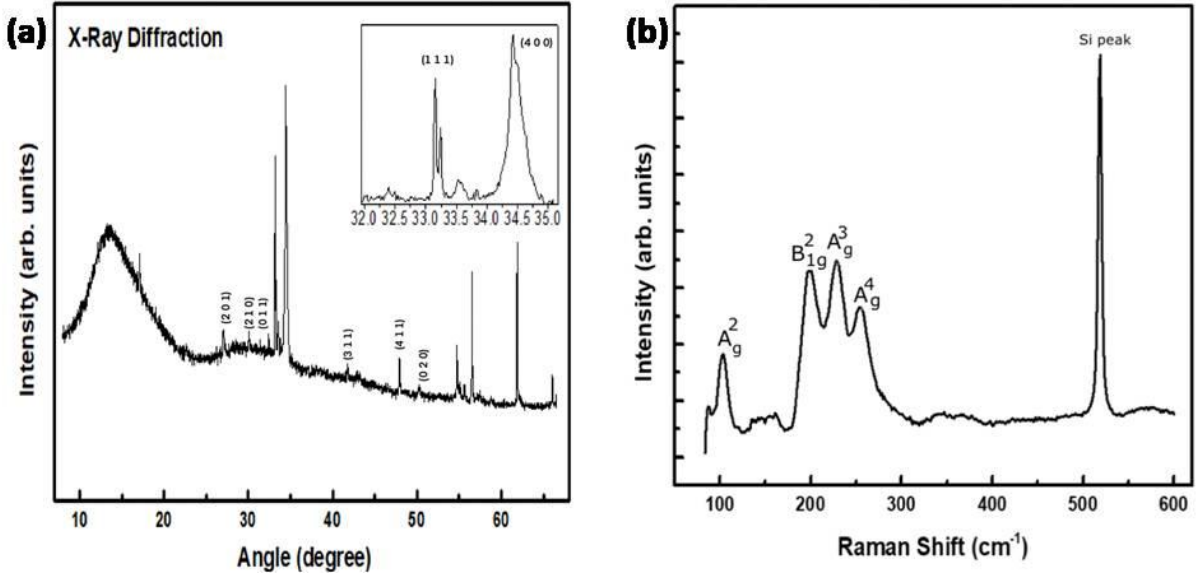

Figure 2: PXRD pattern and Raman spectra of twisted GeS nanowires. (a) PXRD pattern showing peaks which exactly corresponds to the already reported PXRD peaks of GeS. (b) Raman scattering spectra of twisted GeS nanowires with different phonon modes.

The elemental compositions of the nanowire was determined by EDS which gives stoichiometric ratio of Ge:S~1:1. We found that these nanowires were very stable over time both in chemical composition and as well as their structure. To confirm this we repeated the XRD and EDS measurements on the same nanowires after 6 months synthesis. We obtained exactly same XRD and EDS results. The stability of LPCVD grown GeS is considered to be due to the self-encapsulation in a protective, thermally and chemically stable $GeS_X$ shell during synthesis.[16,17]

To measure current-voltage characteristics and photoresponse, devices were fabricated on single GeS nanowires (for more details see materials and methods section). The source drain current vs source drain voltage ($I_{ds}$ vs $V_{ds}$) curve of a single nanowire was measured at three different illumination conditions (fig.3a) in a home-built probe station at room temperature. In dark $I_{ds}$ vs $V_{ds}$ data was almost flat compared to that in 535 nm light illumination. Under white light illumination $I_{ds}$ vs. $V_{ds}$ characteristics was more steeper than that of 535 nm light illumination. The GeS channel and Au electrode contact changes from Schottky to ohmic contact with the light illumination, which is due to photoinduced reduction of Schottky barrier, similar to recently reported studies on the GeS photodetector.[18] The source drain current vs. source drain voltage ($I_{ds}$ vs. $V_{ds}$) (fig.3c) were measured at different gate voltages ($V_g$ = -50 to +50 V), this shows that with increasing negative back gate bias source drain current is increasing considerably because of introduction of more carriers inside the channel. The source-drain vs gate voltage ($I_{ds}$ vs $V_g$) curve at $V_g = +20V$ (fig.3b) shows the p-type behaviour of GeS-FET beacuse of Ge vacancies . The mobility of charge carriers in the GeS-FET can be deduced from $\mu = \frac{L}{WCV_{ds}} \frac{dI_{ds}}{dV_g}$, where L and W are the length and width of the semiconductor channel of the twisted GeS nanowires, respectively, and C is the capacitance between the channel and the back gate per unit area ($C = \frac{\varepsilon_0 \varepsilon_r}{d}$ ; $\varepsilon_0$ is the vacuum permittivity, $\varepsilon_r$ is the relative permittivity, and d is the thickness of the SiO$_2$ dielectric layer).[15] The field-effect mobility of the device is calculated to be 0.70 cm$^2$V$^{-1}$s$^{-1}$; furthermore, the GeS-FET exhibits a on/off ratio of 3 (fig. 3b).This mobility of charge carriers in GeS twisted nanowires is larger than the reported values of mobility in GeS flakes.[12,15,19] However, measured mobility for twisted GeS nanowires is low compared to reported values for other 2D materials[12,15,19] because of large number of defects in the as grown twisted GeS nanowire . These defects causes local variation in the conduction and valence band[20] and the holes gets localized in the forbidden energy gap,[21] which results in a low electrical mobility of the charge carriers. These defect states trap the photogenerated carriers and increase

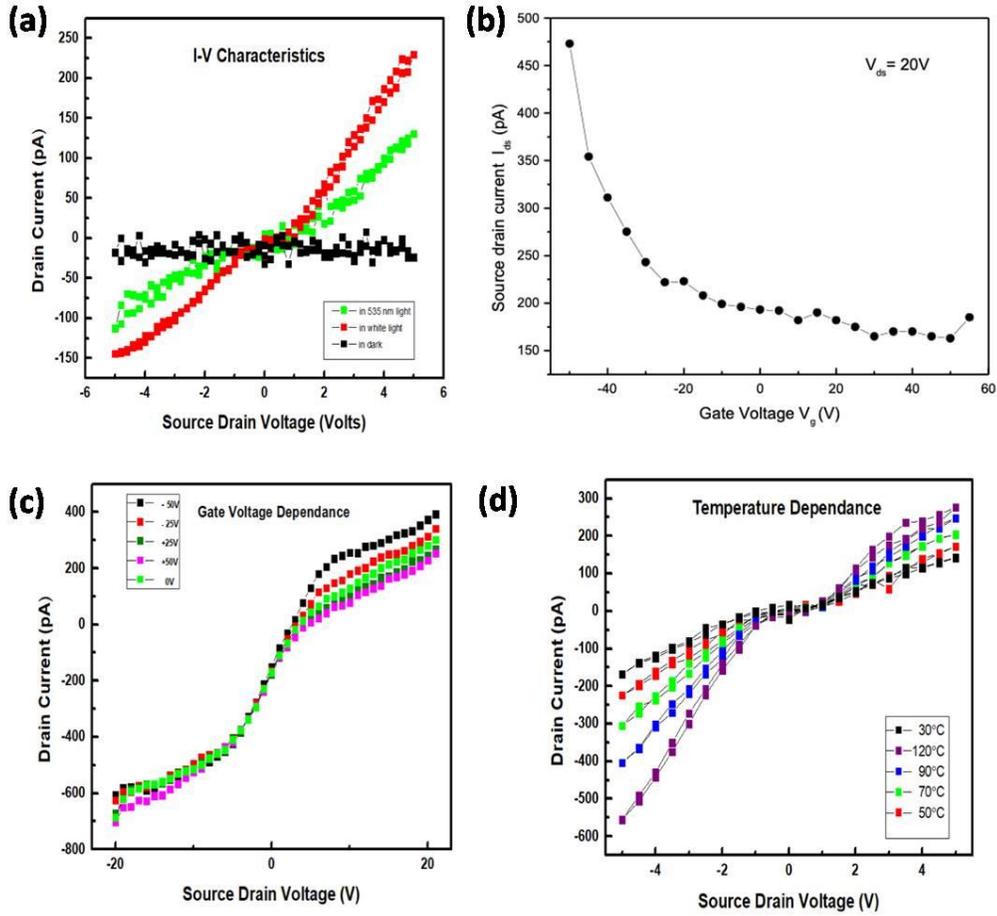

Figure 3: (a) I-V characteristics of twisted GeS nanowires in three different light illumination condition. (b) The source-drain vs gate voltage curve at $V_g$= +20V. (c) Gate voltage dependence of source drain current in twisted GeS nanowires in the range $V_g$= +50V to -50 V. (d) Temperature dependence of source drain current in twisted GeS nanowires in the range (30- 120°C)

the quantum efficiency of GeS due to the longer lifetime of the charge carriers at defect states.[18] To study the effect of thermally generated carriers we measured the source drain current vs. source drain voltage ($I_{ds}$ vs. $V_{ds}$) curve (fig. 3d) at different temperatures (30- 120°C). Here we observed that with increase of temperature, the $I_{ds}$ vs. $V_{ds}$ becomes more and more steeper and the asymmetry of the $I_{ds}$ vs. $V_{ds}$ curve with positive and negative bias also increases. The decrease in resistance can be attributed to the onset of dominating nearest neighbour hopping conduction[22] with increase of temperature.[15] The asymmetry in $I_{ds}$ vs. $V_{ds}$ curve can be attributed to two reasons (i) drift due to a net electric field along the wire axis and (ii) difference in carrier mobility and diffusion coefficient due to charge impurity scattering.[11]

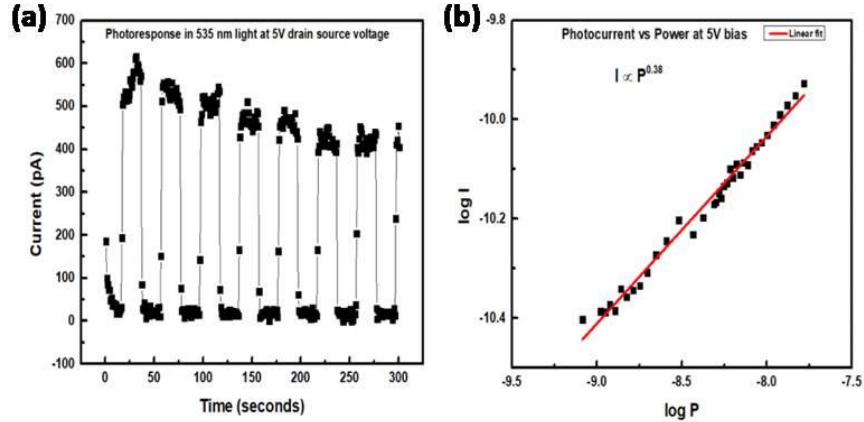

Figure 4: a. Time-resolved photoresponse of the GeS nanowires b. dependence of photocurrent on optical-power (log-log scale)

Earlier we saw that with light illumination, current in the nanowire was observed to increase considerably with respect to dark state. We studied the time-resolved photoresponse (fig. 4a) of the GeS nanowires with a train of on-off illumination of 535 nm wavelength LED at $V_g$=0V and $V_{ds}$ = +5V over a period of 5 minutes and an on-off sequence of 20 secs. The reproducible on-off shapes and a good rise time demonstrate reproducible response and switching stability of the twisted nanowires. The source drain current vs time ($I_{ds}$ vs t) curve shows that the on current decreases with time; this indicates the presence of trap states. To know more about the trap states we studied the intensity dependant photocurrent ($I_{ph}$) measurement by continuously varying the intensity of the 535 nm LED at $V_{ds}$=+5V. The intensities were converted to power (P) and log $I_{ph}$ vs. log P was plotted (fig. 4b), and from the slope of the plot we got the power law to be $I_{ph} \propto P^{0.38}$. This sublinear dependence of photocurrent on optical power indicates presence of trap states and trap limited conduction in GeS.

It has been theoretically predicted that GeS will show intrinsic ferroelectricity at room temperature because of its high critical temperature ($T_C$) although it is a centrosymmetric molecule,[3,7–9] but there is no experimental proof of the same as of now. We drop casted the nanowires on an indium tin oxide (ITO) coated glass substrate[23] and carried out piezoresponse force microscopy (PFM)[24] on these twisted nanowires to check if they have spontaneous polarization state and switching behaviour under an externally applied electric field. A hysteric behaviour of the nanowire can be observed (fig 5a, b) in the "off-state" piezoresponse signal at room temperature. This hysteric behaviour with a sharp 180° switching of electric polarization at different applied dc bias ($V_{dc}$) in fig. 5a can be clearly seen. A large coercive field ~ +10V indicates a strong energy barrier between two opposite polarization states. The domain's polarization direction is dictated by phase of PFM response signal, and for a ferrolectric sample, phase of PFM signal should show hysterisis with 180° polarization switching.[24] So the hysteris curve (fig.5a) shows that the twisted GeS exhibits intrinsic ferroelectricity. The amplitude of PFM response signal is related to the magnitude of the local electromechanical response experienced by the PFM cantilever[24] and these twisted GeS nanowires showed butterfly shaped hysterisis loop (fig.5b) which also indicates the presence of intrinsic ferroelectricty in these twisted GeS nanowires. This evidence of intrinsic ferroelectricity in a centrosymmetric molecule like GeS though counterintutive but has already been theoretically predicted.[7–9] If the inversion symmetry is spontaneously broken then GeS can show intrinsic ferroelectricity. In this case the Eshelby twist in these LPCVD grown GeS nanowires is breaking the inversion symmetry and making the twisted GeS nanowires intrinsically ferroelectric.

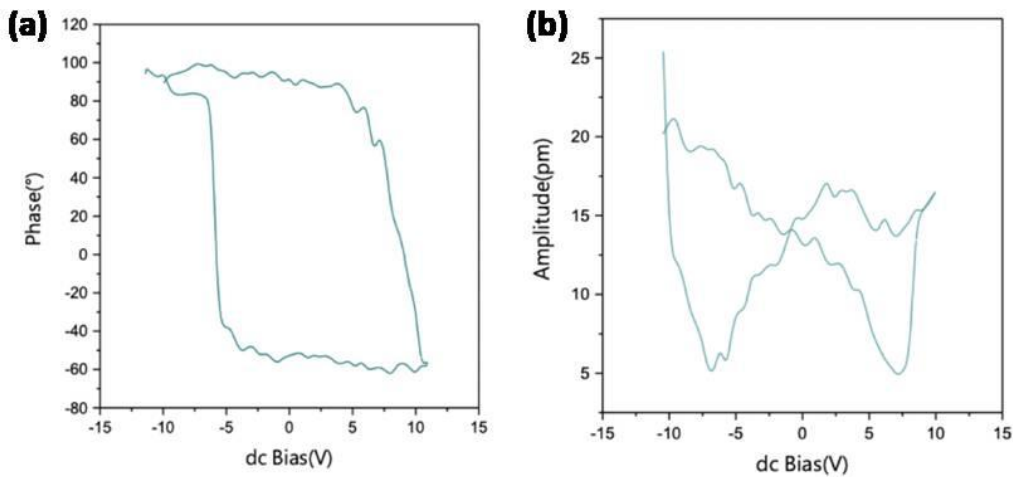

Figure 5: (a) Phase and (b) amplitude of SS-PFM signal from twisted GeS nanowire at room temperature exhibiting ferroelectricity with 180° polarization switching.

Conclusion

In summary, we demonstrated growth of twisted GeS nanowires using low pressure chemical vapor deposition. Raman scattering, PXRD, SAED and EDS were done on these twisted nanowires to confirm the chemical composition and crystallinity. The twist in these nanowires can be easily seen in FE-SEM. Also the crystallographic twist is confirmed by TEM and SAED. Transfer characteristics of these nanowires showed that the twisted nanowires were p-type. The field-effect mobility of the device is calculated to be 0.70 $cm^2V^{-1}s^{-1}$ which is highest reported so far for GeS. Source drain current was strongly dependant on back gate voltage and on light illumination. Transient photoresponse results showed that these nanowires have sustainable and robust photoswitching property. Intensity dependant photocurrent study shows the presence of trap states either inside the channel of nanowire or presence of trapped impurities in the interface of GeS and $SiO_2$ dielectric. Finally PFM measurement showed that these GeS nanowires have intrinsic ferroelectric order though GeS is a centrosymmetric molecule. This appearance of ferroelectricity is due to spontaneously broken centrosymmetry because of Eshelby twist in these LPCVD grown GeS nanowires. This emergence of room temperature ferroelectricity in the semiconducting GeS twisted nanowires is an important advancement for the electronic and optoelectronic devices.

Materials and Methods

Twisted GeS nanowires were synthesized using GeS powder (99.99%, Sigma Aldrich) in an experimental setup consisting of a pumped quartz tube with two temperature zone. The evaporation zone containing a quartz boat with the GeS powder was heated to 450°C, while the zone containing the substrate was heated to growth temperatures of 270−300°C. Si (100) covered with 3 nm thick Au films deposited by sputtering at room temperature was used as substrate. Such Au films dewet at the growth temperature. During growth Ar gas buffered with $H_2$ (2%) carrier gas flow was maintained at 50 standard cubic centimeters per minute (sccm) at a pressure of 1-2 Torr. To measure current-voltage characteristics and photoresponse, devices were fabricated on single GeS nanowires. First, from Si substrates the nanowires were scraped out and dispersed in 99% ethyl alcohol solution. The nanowires were then drop casted on 300 nm $SiO_2$ coated Si substrate and a two probe contact of Cr/Au (5/65 nm) was made on these nanowires by standard photolithographic process. All the electrical characterisations were done in a homebuilt probe station with the help of Keithley 2450 source measure unit and Yokogawa GS820 multi channel source measure unit. FE-SEM images were captured through ZEISS Ultra Plus and TEM images were captured using a JEOL JEM-2200FS transmission electron microscope. EDX compositional analysis and elemental mapping were carried in TEM imaging mode. XRD data were recorded at room temperature from a Bruker D8 Advance diffractometer using Cu $K_\alpha$ radiation (λ = 1.5406 Å). Raman spectra (λ = 633 nm) were recorded at Raman microscope (LabRAM HR, Horiba Jobin Yvon) with a 60X objective lens. PFM measurements were done using an Asylum research AFM (MFP–3D) with an additional high voltage amplifier. The sample was mounted on a conducting sample holder which was directly connected to the ground of the amplifier. The conductive AFM cantilever having a Pt-Ir tip on it was brought in contact with the sample (nanowires). An AC excitation of 4V riding on a dc bias voltage ($V_{dc}$) was applied between the tip and the amplifier ground. The response of the sample to the electrical stimulus was detected through the reflection of the laser beam from the end of the cantilever onto a position sensitive photo diode. In order to ensure that the hysteretic effects are due to ferroelectricity which may otherwise arise from electrostatic and electrochemical effect, all the measurements were performed following SS-PFM (switching spectroscopy piezoresponse force microscopy) initiated by Jesse et al.[24–26] In this method, instead of sweeping $V_{dc}$ continuously, bias is applied in sequence of pulses where the phase and amplitude measurements are done in the "off" states and an appreciable change is observed in the "off-state" results as compared to the "on-state" measurements which is a clear evidence of the minimization of electrostatic effects. We further performed the topographic imaging after the spectroscopic measurements, where no topographic modification was observed which usually occurs due to the electrochemical reaction between the tip and sample.[24]

Acknowledgement: TC acknowledges CSIR, Government of India for JRF fellowship (File no. **09/936(0280)/2019-EMR-I**). GS acknowledges financial support from the Department of Science and Technology (DST) through Swarnajayanti fellowship (Grant number: **DST/SJF/PSA-01/2015-16**). The authors acknowledge Ms. Sandra Sajan, for her help in PFM characterizations